\documentclass{ws-procs975x65}

\usepackage{amssymb}
\usepackage{amsmath}
\usepackage{amsfonts}
\usepackage{graphicx}

\newcommand\beq{\begin{equation}}
\newcommand\eeq{\end{equation}}
\newcommand\bea{\begin{eqnarray}}
\newcommand\eea{\end{eqnarray}}
\newcommand\bseq{\begin{subequations}} 
\newcommand\eseq{\end{subequations}}
\newcommand\bal{\begin{align}}  
\newcommand\ealign{\end{align}}    

\begin{document}

\title{Mixmaster Chaos via the 
Invariant Measure}
			
\author{GIOVANNI IMPONENTE}

\address{Dipartimento di Fisica, 
 						Universit\'a ``Federico II'', 
 						Napoli  and 
					 INFN -- Napoli -- Italy \\
					ICRA -- International Center for
								Relativistic Astrophysics \\
E-mail: imponente@icra.it}

\author{GIOVANNI MONTANI}

\address{Dipartimento di Fisica -- G9
					Universit\'a ``La Sapienza'', Roma -- Italy \\
 ICRA -- International Center for
								Relativistic Astrophysics \\
				E-mail: montani@icra.it}  


\maketitle

\abstracts{
The chaoticity of the Mixmaster is discussed in 
the framework of Statistical Mechanics by using 
Misner--Chitre-like variables and an ADM reduction 
of its dynamics. \\
We show that such a system 
is well described by a microcanonical ensemble 
whose invariant measure is induced by the corresponding 
Liouville one and is uniform.
The covariance with respect to the choice of the 
temporal gauge  of the obtained invariant measure 
is outlined.
}

\section{Introduction}

The original treatment of the Mixmaster \cite{M69} 
chaoticity due to Belinski, Khalatnikov and 
Lifshitz (BKL) \cite{BKL70} provided a 
satisfactory description of the Kasner indices 
behaviour, especially because the invariant measure 
for the Poincare return map was calculated in terms 
of the $u$ parameter. Nevertheless the question 
about how to construct a presentation for the 
system stochasticity in terms of continuous 
variables remained open.
Relevant achievements in this direction appeared 
in \cite{CB83,KM97,IM02}, where an invariant 
measure associated to the motion of the billiard 
ball representing the system is studied in terms 
of continuous Misner--Chitre-like variables.

Here we review the fundamental steps at the 
grounds of this analysis and focus our attention 
to the invariance of the statistical mechanics 
representation with respect to the choice of 
the lapse function \cite{IM02}.

\section{Asymptotic Dynamics} 
 
Near the cosmological singularity the Mixmaster 
model admits a dynamics described by the 
two-dimensional ADM reduced action 
\begin{equation} 
 \mathcal{S}_{\textrm{RED}}=\int_{\Gamma_H} 
 \left( p_{\xi}d \xi  +  p_{\theta} {d\theta} -  
 \varepsilon {df} \right)\, ,
\label{q} 
\end{equation} 
being 
\beq
\varepsilon ^2 = \left({\xi}^2 -1\right){p_{\xi}}^2 
+\frac{{p_{\theta}}^2}{{\xi}^2 -1} \, ;
\label{d2}
\eeq
we adopted Misner--Chitre-like variables
$\xi, \theta, f(\tau)$ \cite{IM01} and 
$p_{\xi}$ and $p_{\theta}$ denote the conjugate 
momenta to $\xi$ and $\theta$, respectively. \\
This picture is associated to the temporal 
gauge 
\begin{equation} 
N\left(\tau\right)= \frac{12 D}{E} 
e^{2f}  \frac{df}{d\tau}  \, ,
\label{rs} 
\end{equation} 
in which 
$D= \exp( -3 \xi e^{f(\tau)} )$
and $E$ denotes a generic positive 
value for the energy-like constant of 
motion $\varepsilon$ and the fixing of  
a specific time variable 
corresponds to choose a suitable function 
$f (\tau)$. The integral in (\ref{q}) is 
referred to the domain $\Gamma_H$ 
(see Figure 1
) outside of which 
the motion is classically forbidden; 
$\Gamma_H$ is dynamically closed and 
corresponds to a portion of a two-dimensional 
Lobachevsky plane, i.e. a surface of constant 
negative curvature.
\begin{figure}
\label{fig:cusp}%
\centering
\includegraphics[width=0.7\hsize]{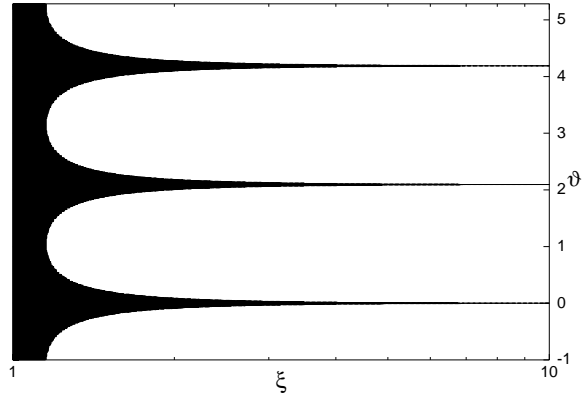}
\caption{Domain $\Gamma_H$ dynamically accessible}
\end{figure}

\section{Invariant Measure}

This representation of the Mixmaster dynamics in 
terms of a billiard ball on a Lobachevsky plane 
is manifestly chaotic and the existence of an 
energy-like constant of motion characterizes the 
statistical properties 
by a \textit{microcanonical ensemble};
therefore the distribution function takes
the Liouville form
\beq
d\mu \propto \delta(E-\varepsilon)d\xi d\theta
dp_{\xi} dp_{\theta} \, .
\eeq
Since the value of $\varepsilon$ cannot contain 
any information about the Mixmaster chaoticity
then we have to integrate on this variable 
by redefining the 
momenta as follows
\bseq
\begin{align} 
p_{\xi } &= \frac{\varepsilon}{\sqrt{\xi ^2 - 1}}\cos\phi  \\ 
p_{\theta } &= \varepsilon \sqrt{\xi ^2 - 1}\sin\phi ,  
\label{v} 
\end{align} 
\eseq
where $0 \leq \phi < 2\pi$;
hence we get an uniform invariant measure as written 
in the variables $\xi, \theta, \phi$
(in the phase space $\varepsilon$ assumes a fixed 
value), i.e.
\begin{equation} 
w_{\infty} =  \left\{ \begin{array}{l}
					\displaystyle	\frac{1}{8\pi ^2} \, \qquad \qquad\textrm{in~}\Gamma_H \\
								0							\, \qquad\qquad\textrm{outside~}\Gamma_H
									\end{array}		
											\right.
\label{x} 
\end{equation} 
The invariant measure (\ref{x}) holds in the time 
variable $f$ and can be taken as the stationary solution 
of the Liouville equation 
\begin{align}
\label{xx}
\frac{dw}{df} =
	\frac{\partial w}{\partial f} 
	+ \frac{d \xi}{df} \frac{\partial w}{\partial\xi} 
	&+\frac{d \theta}{df} \frac{\partial w}{\partial\theta}
	+\frac{d \phi}{df} \frac{\partial w}{\partial\phi} = \nonumber \\
&= \frac{\partial w}{\partial f} +\sqrt{\xi^2-1}\cos\phi \frac{\partial w}{\partial\xi} 
 +\frac{\sin\phi}{\sqrt{\xi^2-1}}\frac{\partial w}{\partial\theta} 
-\frac{\xi\sin\phi}{\sqrt{\xi^2-1}}  \frac{\partial w}{\partial\phi} =0 \, ,
\end{align}
associated to the Hamiltonian system
\begin{align}
\frac{d\xi}{df}=\sqrt{\xi^2-1}\cos\phi  \, , \quad
\frac{d\theta}{df}=\frac{\sin\phi}{\sqrt{\xi^2-1}}  \, , \quad
\frac{d\phi}{df}=-\frac{\xi\sin\phi}{\sqrt{\xi^2-1}} \, .
\end{align}
The invariance of distribution (\ref{x}) is ensured 
by the holding of Eq. (\ref{xx}) in any other time 
variable; in fact, the time gauge (\ref{rs})
allows to re-express the Liouville theorem (\ref{xx})
in a generic variable $\tau$ via a simple multiplication 
by a common factor. \\
Indeed for our proof is relevant that we are dealing 
with a stationary distribution function; for 
non-stationary corrections to the measure (\ref{x})
and the corresponding 
asymptotic behaviour of the right-hand side 
of (\ref{xx}) see \cite{M01}.
However, all such non-stationary terms decay 
exponentially and do not affect the results here 
presented.

\end{document}